\documentstyle[twocolumn,floats,epsf,prl,aps]{revtex}

\begin{document}
\draft
\preprint{HEP/123-qed}
\title{$\bf \hat{c}$-Axis Electrodynamics of $\bf YBa_2Cu_3O_{7-\delta}$}
\author{A. Hosseini, Saeid Kamal, D.A. Bonn, Ruixing Liang, and W.N. Hardy}
\address{Dept. of Physics and Astronomy, University of British Columbia,
6224 Agricultural Rd., Vancouver, BC, V6T 1Z1, Canada}
\date{\today}
\maketitle
\begin{abstract}
New measurements of surface
impedance in $YBa_2Cu_3O_{7-\delta}$ 
show that
the $\hat{c}$-axis
 penetration depth and conductivity below $T_c$
exhibit behaviour different from that observed
in the planes. The $\hat{c}$-axis penetration depth never has the linear
temperature dependence seen in the ab-plane. Instead of the conductivity
peak seen in the planes, the $\hat{c}$-axis
microwave conductivity falls
to low values in the superconducting state, then rises
slightly below 20~K.  
These results show that $\hat{c}$-axis transport
remains incoherent below $T_c$, 
even though this is one of the least anisotropic cuprate 
superconductors.

\end{abstract}

\narrowtext

The highly anisotropic nature of cuprate superconductors 
and the question of their dimensionality continues to play a central
role in research on high temperature superconductors. 
Early
on it was established experimentally that transport properties 
within the $CuO_2$ planes of these materials differ markedly from
transport normal to the planes.
The DC resistivity in the $\hat{c}$ direction
can be orders of magnitude higher than it is within the planes and it 
often has a qualitatively different temperature dependence, exhibiting
upturns with decreasing temperature \cite{cooper}.
The 2 dimensional character
that this normal state anisotropy implies has been an essential
ingredient in many of the ideas
that have been brought forward to explain the unusual properties 
and the high critical temperatures of these materials \cite{anderson}.
However, $YBa_2Cu_3O_{7-\delta}$,
which is one of the best controlled
and most studied of the high temperature superconductors, is also one 
of the least anisotropic members of the family. For low
oxygen vacancy levels
$\delta < 0.07$ the c-axis resistivity shows no upturn with decreasing 
temperature and the resistivity anisotropy is less than a factor of 50
\cite{friedman}. In this letter we present new measurements of the
c-axis microwave conductivity and penetration depth that clearly show
that in the superconducting state, the interplane electrodynamics
are quite different from those observed in the ab-plane. 
Both quantities show that transport in the $\hat{c}$-direction is
incoherent, even though the anisotropy in the normal state is not 
particularly large.

One of the strengths of measurements of the surface impedance 
$Z_s=R_s+iX_s$
is that they provide complementary information on both the superfluid
density and the low energy excitations out of the condensate. In the 
limit of local electrodynamics, the surface reactance 
$X_s(T)=\mu_0\omega\lambda(T)$
provides a very
direct measurement of the London penetration depth $\lambda(T)$, which
in turn gives the superfluid density $n_s(T)$ via 
$\lambda^{-2}(T)=\mu_0n_s(T)e^2/m^*$. The
surface resistance in the superconducting state
is given by \cite{equa}  
$R_s=\mu_0^2\omega^2\lambda^3(T)\sigma_1(\omega,T)/2$, thus
providing a means
of observing what happens to the low frequency
conductivity $\sigma_1(\omega,T)$
of the material as it is
cooled below the superconducting transition.
  
The ab-plane surface impedance of $YBa_2Cu_3O_{7-\delta}$ has been
extensively studied, revealing several key features of the 
superconducting state. Measurements of $\lambda(T)$ at low T in 
$YBa_2Cu_3O_{7-\delta}$ exhibit a linear temperature dependence
\cite{hardy93},
which is a strong indication of nodes in the energy gap of this
material. Closer to $T_c$, $\lambda(T)$ has been found to vary as 
$((T_c-T)/T_c)^{-1/3}$,
consistent with a superconducting transition governed
by 3D XY-like critical fluctuations \cite{kamal94}.
The temperature 
dependence of the ab-plane surface resistance reveals an enormous
peak in $\sigma_1(T)$ that has been attributed to a rapid decrease in the
in-plane scattering rate below $T_c$ \cite{bonn92}.

Obtaining this kind of information for currents running in the
$\hat{c}$-direction presents great technical difficulties. 
Most surface impedance measurements involve placing a sample
in microwave magnetic fields $\vec{H}_{AC}$,
where the currents induced in the surface must form
closed loops. By changing the geometry or orientation of the sample
one can perform measurements that contain different admixtures of ab-plane
and $\hat{c}$-axis currents and then extract the surface
impedance in different directions.
This can be achieved by rotating a sample 
or by measuring samples cut with different orientations. 
Measurements that involve rotating
the sample \cite{kitano,mao} 
have severe problems due to changing demagnetizing factors and changing
current distributions. 
If one is working with thin,
rectangular samples, which is the most common situation with crystals
of cuprate superconductors, there is a huge change in demagnetizing 
factors if the sample is rotated from $\vec{H}_{AC}\parallel\hat{c}$
(planar currents only)
to $\vec{H}_{AC}\perp\hat{c}$ (combination of planar and 
$\hat{c}$-axis currents).
Even if the sample is not very thin and the demagnetizing factors can
be dealt with, changes in the ab-plane
current distribution bring poorly controlled uncertainties into such
procedures for
extracting the $\hat{c}$-axis electrodynamics.

For $La_{1.85}Sr_{0.15}CuO_4$, where large samples are available,
Shibauchi et al. arrived at a more satisfactory solution
by cutting and polishing
thin slabs with different orientations \cite{shibauchi94}. This avoids the
problem of changing demagnetizing factors, although it does force one to
compare measurements on different samples, thus relying on two slabs
being otherwise identical. 
Our approach to obtaining the c-axis surface impedance relies on the
fact that thin crystals of $YBa_2Cu_3O_{7-\delta}$ cleave very cleanly
in the [100] and [010] directions. The surface resistance or 
penetration depth is initially measured with the microwave magnetic field 
lying in the plane of the thin slab, measuring with
both $\vec{H}_{AC}\parallel \hat{b}$ ($\hat{a},\hat{c}$ currents) and
$\vec{H}_{AC}\parallel \hat{a}$ ($\hat{b},\hat{c}$ currents).
The contribution due to $\hat{c}$-axis currents is then increased
by cleaving the slab into a set of
narrow needles which is remeasured
with $\vec{H}_{AC}$ lying along the axis of the needles. 
This technique is particularly reliable because it has no significant
change in demagnetizing factors, no change in the distribution of ab-plane
currents, and there is
no need to compare different samples. With this sequence of experiments
and measurements of the sample's dimensions it is straightforward to
extract the surface impedance in all three directions.
We have measured $\lambda(T)$ in this way in a 
superconducting loop-gap resonator and the results
have been described elsewhere
\cite{bonn96}. 
The new measurements of surface resistance in the
$\hat{c}$-direction $R_{sc}(T)$
have been performed
in a superconducting 22 GHz cylindrical cavity operated in the
$TE_{011}$ mode. The samples are crystals of $YBa_2Cu_3O_{7-\delta}$ 
grown in yttria-stabilized zirconia crucibles, then detwinned and 
annealed to set the oxygen content \cite{liang}.

\begin{figure}
\begin{center}
\leavevmode
\epsfxsize=3in 
\epsfysize=3in 
\epsfbox{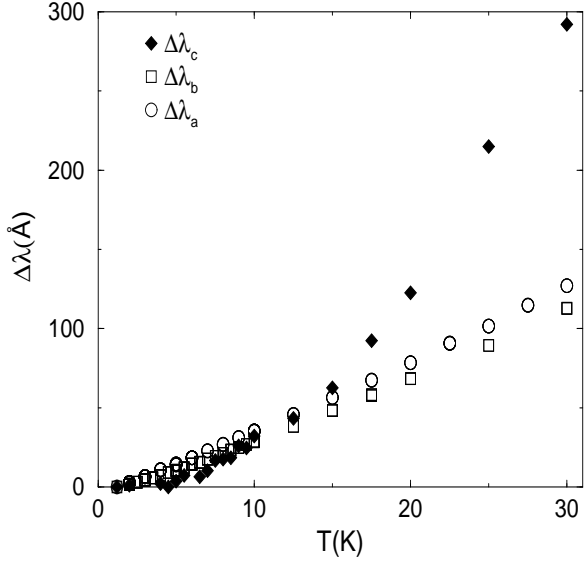}
\end{center}
\caption{The temperature dependence of the penetration depth
$\Delta\lambda(T)=\lambda(T)-\lambda(1.2 K)$ of $YBa_2Cu_3O_{6.95}$
is nearly quadratic in the $\hat{c}$ direction and linear in the ab-plane.}
\label{fig:deltalam}
\end{figure}

Fig.~\ref{fig:deltalam} shows the temperature dependence of the penetration
depth $\Delta\lambda_c(T)=\lambda_c(T)-\lambda_c(1.2 K)$
extracted in the manner described above.
Instead of the linear behaviour seen in both
directions in the ab-plane, the temperature dependence is close to 
quadratic,  
with a power law of 
$\Delta\lambda(T)\propto T^{2.1}$
giving a good fit to the data up to about 40 K. These microwave 
results can
be combined with far infrared measurements of 
$\lambda_c(0)$ \cite{homes,basov}
to produce
\begin{figure}
\begin{center}
\leavevmode
\epsfxsize=3in 
\epsfysize=2.98in 
\epsfbox{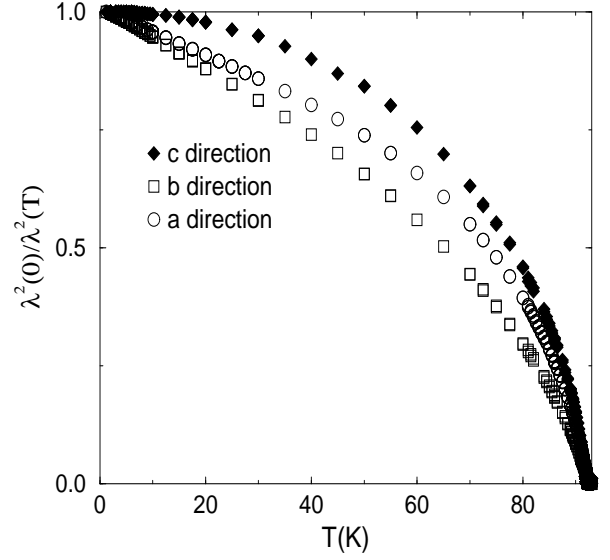}
\end{center}
\caption{The superfluid fraction in the $\hat{c}$ direction of
$YBa_2Cu_3O_{6.95}$ is qualitatively
different from the behaviour seen in either direction in the ab-plane.}
\label{fig:superfl}
\end{figure}
the superfluid fractions shown in Fig.~\ref{fig:superfl}. A key
feature of this figure is that in all three directions, the
behaviour near $T_c$ is
consistent with 3D XY-like critical fluctuations. However, at lower
temperatures, the superfluid fraction in the $\hat{c}$-direction
is very flat and shows no sign of the linear temperature
dependence observed in the ab-plane. Furthermore, previous work has shown
that this behaviour persists over a wide doping range, from underdoped
($\delta=0.42$) to slightly overdoped ($\delta=0.01$) \cite{bonn96}.
This indicates that,
despite the 3D XY critical behaviour near $T_c$,
highly doped $YBa_2Cu_3O_{7-\delta}$ does
not behave as if it were a d-wave
pairing state in an anisotropic 3 dimensional metal,
where one would expect the superfluid density to be linear at low
temperatures in all three directions \cite{radtke96,xiang96a}. 
Strikingly similar behaviour has been reported
by Shibauchi et al. for polished slabs of
$La_{1.85}Sr_{0.15}CuO_4$ \cite{shibauchi94}, by Jacobs et al. for 
cleaved crystals of $Bi_2Sr_2CaCu_2O_{8+\delta}$ \cite{jacobs95}, and by
Panagopoulos et al. for aligned powders of $HgBa_2Ca_2Cu_3O_{8+\delta}$
\cite{panagopoulos96}.

\begin{figure}
\begin{center}
\leavevmode
\epsfxsize=3in 
\epsfysize=3in 
\epsfbox{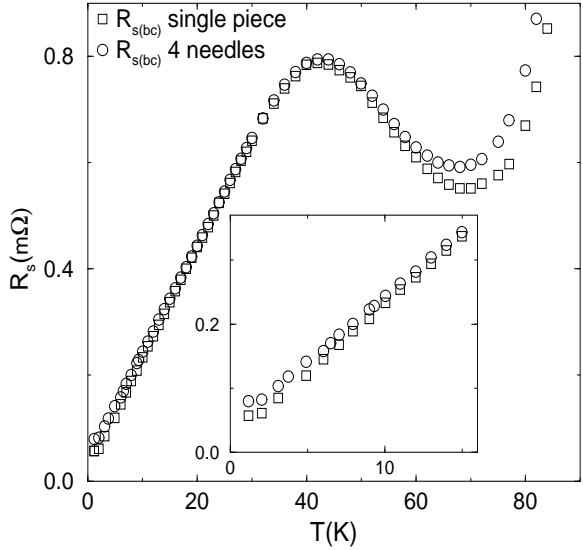}
\end{center}
\caption{The surface resistance at 22 GHz of a thin plate of 
$YBa_2Cu_3O_{6.95}$, before and after cleaving the plate into 4 needles.
The small differences are used to extract the
$\hat{c}$-axis surface resistance
shown in Fig.~\ref{fig:rsabc}.}
\label{fig:rsraw}
\end{figure}
\begin{figure}[t]
\begin{center}
\leavevmode
\epsfxsize=3in 
\epsfysize=3in 
\epsfbox{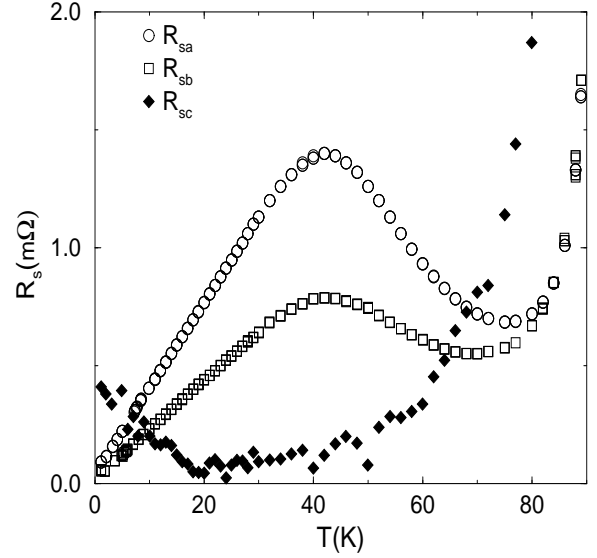}
\end{center}
\caption{The broad peak in $R_{sa}(T)$ and $R_{sb}(T)$ is due to a rapid
increase in quasiparticle lifetime below $T_c$. The surface resistance 
in the $\hat{c}$ direction is 
different from these, showing an upturn only at low temperatures.
}
\label{fig:rsabc}
\end{figure}
Fig.~\ref{fig:rsraw} shows
measurements of $R_s(T)$ at 22 GHz
performed on a thin plate with
$\vec{H}_{AC}\parallel \hat{a}$.
This orientation
produces currents running across the face of the slab in the
$\hat{b}$-direction,
with a small contribution from currents running down the side of the slab in
the $\hat{c}$-direction. The small change seen in $R_s(T)$ after cleaving the
sample into four needles is the increase in loss due to an increased
$\hat{c}$-axis contribution.
Except near $T_c$, the change is extremely small which
indicates that the surface resistance in the $\hat{c}$ direction $R_{sc}(T)$
is actually quite low. Since 
$R_s\propto \lambda^3(T)\sigma_1(\omega,T)$, we see
that the increase in loss that
might be expected from the much larger $\lambda$ in the c-direction is
in fact 
balanced by a rather low c-axis conductivity. The influence of
$R_{sc}(T)$ is most clearly discernable above 60~K, and rather
surprisingly there is some additional c-axis loss appearing below 20~K.
The small size of the effect observed here indicates that it would be
difficult
to unambiguously extract $R_{sc}(T)$ from the earlier techniques 
that involved 
changing the orientation of the sample \cite{kitano,mao}, since 
the effect of changes in current distribution can
easily be larger than the change observed in this experiment.

Fig.~\ref{fig:rsabc} shows the surface resistance in all three directions
($R_{sa}$, $R_{sb}$, $R_{sc}$), extracted
from the data in Fig.~\ref{fig:rsraw}, plus a set
of measurements to
determine $R_{sa}(T)$. $R_{sc}(T)$ is very
low and qualitatively different from that observed in either of the planar
directions. In the ab-plane, $R_s(T)$ in high purity crystals exhibits
a broad peak
which is caused by a very large peak in $\sigma_1(T)$ in both the $\hat{a}$
and $\hat{b}$
directions. This increase in the ab-plane conductivity below
$T_c$ has been attributed to a rapid increase in quasiparticle lifetime 
in the superconducting state \cite{bonn92}, 
but the increase seems to be completely
absent for
carriers moving in the $\hat{c}$-direction.
Instead, $R_{sc}(T)$ falls to very low
values below $T_c$ and then rises slightly again below 20~K.

\begin{figure}
\begin{center}
\leavevmode
\epsfxsize=3in 
\epsfysize=3in 
\epsfbox{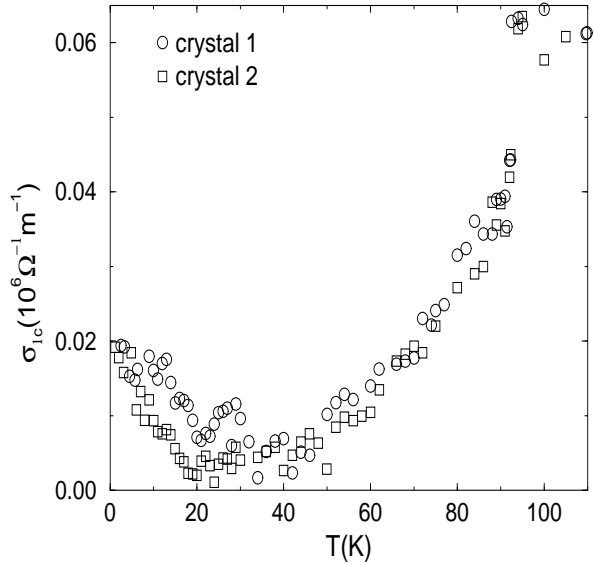}
\end{center}
\caption{Measurements of $\lambda_c(T)$ and $R_{sc}(T)$ 
are used to extract the $\hat{c}$-axis conductivity shown here.
The 
$R_{sc}(T)$ measurements
have been repeated on a crystal taken from a different growth
run.
Although the conductivity is small and difficult to measure it is clearly
reproducible from run to run. Below $T_c$ $\sigma_{1c}(T)$ falls orders
of magnitude below the ab-plane conductivity, then rises again
slightly at low temperatures.}
\label{fig:sigmac}
\end{figure}

Using the measurements of $\lambda_c(T)$, the
c-axis conductivity $\sigma_{1c}(T)$
can be extracted from this measurement of $R_{sc}(T)$.
Because $R_{sc}(T)$
is so small, and rather surprising 
in shape, we have repeated the entire set of 
measurements on a sample taken from a different crystal growth run
and the conductivity from 
both sets of data is shown in Fig.~\ref{fig:sigmac}.
In the normal state just above $T_c$,
the microwave conductivity of both crystals is about
$6.3\times10^4\, \Omega^{-1}m^{-1}$, corresponding
to a DC resistivity of $1.6 \, m\Omega cm$, which is in
good agreement with the range of values 
reported for crystals of $YBa_2Cu_3O_{7-\delta}$
in this range of oxygen doping \cite{friedman}. Below $T_c$, $\sigma_{1c}(T)$
falls rapidly, with no sign of the peak observed in the ab-plane,
the conductivity eventually reaching a minimum value near 30 K.
The qualitatively different temperature
dependencies in the two directions leads to
an anisotropy in the conductivity of almost $10^4$ by 30~K!
However, $\sigma_{1c}(T)$ never falls to zero, but instead rises again at
low temperatures, so the $\hat{c}$-axis never
completely exhibits insulating behaviour.
At 10~K the value of $\sigma_{1c}(T)$ at 22 GHz is 
$10^4\Omega^{-1}m^{-1}$, which is close to the 
residual conductivity of
$\sigma_{1c}(T=10K)\approx 5\times 10^3\Omega^{-1}m^{-1}$
at 100~$cm^{-1}$,
the low frequency limit of far infrared 
$\hat{c}$-axis measurements \cite{homes}.

Both the
magnitude of $\sigma_{1c}(T)$ and its rapid drop in the superconducting
state suggest that $\hat{c}$-axis transport is incoherent,
even below $T_c$ in $YBa_2Cu_3O_{6.95}$. This resolves a conflict presented 
by earlier surface impedance measurements that showed a broad peak in
$\sigma_{1c}(T)$, similar to the one observed in the
ab-plane conductivity \cite{kitano,mao}. The absence of such a peak in the 
data presented here indicates that the $\hat{c}$-axis transport is not
influenced by the development of the
long transport lifetimes seen in 
ab-plane measurements below $T_c$, and is better approached as a
case of incoherent transport.
This is in accord with the conclusion
that the lack of a linear temperature dependence in $\lambda_c(T)$ indicates
that the $\hat{c}$-axis superfluid density is governed by incoherent processes
\cite{radtke96,xiang96a}.

A common approach to treating this incoherent transport is to model it as
Josephson tunneling, where the superfluid density takes the form
$\lambda_c^2(0)/\lambda_c^2(T)\propto \Delta(T)tanh(\Delta(T)/2k_BT)$
\cite{clem}. We find that this expression fits the low temperature data
for $YBa_2Cu_3O_{7-\delta}$ poorly. Instead, a power law, close to $T^2$
gives a better fit from 1 to 40 K and a careful look at the published data
on other systems suggests that this nearly quadratic temperature dependence
is common to many materials. In fact,
$\lambda_c^2(0)/\lambda_c^2(T)$ looks very similar in
$YBa_2Cu_3O_{7-\delta}$ \cite{bonn96}, $Bi_2Sr_2CaCu_2O_{8+\delta}$
\cite{jacobs95},
$La_{1.85}Sr_{0.15}CuO_4$ \cite{shibauchi94}
and $HgBa_2Ca_2Cu_2O_{8+\delta}$ \cite{panagopoulos96}, despite 
substantial structural variations. This temperature dependence seems 
largely independent of wide variation in the degree of anisotropy,
as measured by either the normal state transport anisotropy or the wide
variation in $\lambda_c(0)$ across this set of materials. This
argues against models of layered superconductors
where the degree of anisotropy and structural details are
correlated with the
temperature dependence of the $\hat{c}$-axis superfluid density. 
One possible source of $T^2$ dependence in the $\hat{c}$-axis
superfluid density is impurity assisted hopping 
\cite{radtke96,xiang96b}, but a similar power law also comes from a pair
tunneling model that produces the main features seen
in all three directions in
$YBa_2Cu_3O_{6.95}$ \cite{xiang96a}. Differentiating between various 
models comes up against the central 
issue of whether or not the incoherence in the $\hat{c}$-direction
is an intrinsic feature   
of these systems, exemplified by theories
involving ``confinement''\cite{clarke95},
or is a consequence of weak coupling between quasi-2D layers that nevertheless
behave
as Fermi liquids \cite{radtke96}. The very low $\hat{c}$-axis conductivity
shown here for $YBa_2Cu_3O_{6.95}$ below $T_c$, and the incoherent behaviour
of the $\hat{c}$-axis superfluid density in all cuprate systems studied so far,
provide important tests of these different points of view.

\acknowledgments

We wish to acknowledge helpful conversations with P.J. Hirschfeld, 
I. Affleck, C. Homes, D. Basov, and T. Timusk.
This research was supported by the Natural Science and Engineering 
Research Council of Canada and the Canadian Institute for Advanced
Research. DAB acknowledges support from the Sloan
Foundation.


\begin{references}
\bibitem{equa}Note that this is a useful simplifiation, valid in the local
limit for 
$\sigma_2\gg\sigma_1$, which occurs as soon as $T$ is slightly below
$T_c$.
\end{references}
\end{document}